\documentclass {jaa}
\usepackage[english]{babel} 
\usepackage{graphics}
\usepackage{graphicx}
\usepackage{color}
\usepackage{amsmath}
\usepackage{psfrag}
\usepackage{epsfig}
\usepackage[round]{natbib}

\newcommand{\rsun}{R$_\odot$}
\newcommand{\kmps}{kms$^{-1}$}

\begin{document}

\title[Space Weather and Solar Wind Studies with OWFA] 
{Space Weather and Solar Wind Studies with OWFA}
\author[Manoharan, Subrahmanya, \& Chengalur] 
{P. K. Manoharan$^{1}$\thanks{Email: mano@ncra.tifr.res.in},
 C. R. Subrahmanya$^{2}$\thanks{Email: crs@rri.res.in}, 
 J. N. Chengalur$^{3}$\thanks{Email: chengalur@ncra.tifr.res.in}\\  \\
 $^{1}$ Radio Astronomy Centre, NCRA-TIFR, P.O. Box 8, Udhagamandalam (Ooty) 643001 \\
 $^{2}$ Raman Research Institute, C. V. Raman Avenue, Sadashivnagar, Bengaluru 560080 \\
 $^{3}$ NCRA-TIFR, Pune University Campus, Ganeshkhind, Pune 411007
}
\date {}
\maketitle

\begin{abstract}
In this paper, we review the results of interplanetary scintillation 
(IPS) observations made with the legacy system of the Ooty Radio 
Telescope (ORT) and compare them with the possibilities opened by
the upgraded ORT, the Ooty Wide Field Array (OWFA). The 
stability and the sensitivity of the legacy system of ORT allowed the 
regular monitoring of IPS on a grid of large number of radio sources and 
the results of these studies have been useful to understand the physical 
processes in the heliosphere and space weather events, such as 
coronal mass ejections, interaction regions and their propagation 
effects. In the case of OWFA, its wide bandwidth of 38 MHz, the large 
field of view of $\sim$27$^\circ$ and increased sensitivity provide 
a unique capability for the heliospheric science at 326.5 MHz. IPS 
observations with the OWFA would allow one to monitor more than 5000 
sources per day. This, in turn, will lead to much improved studies of 
space weather 
events and solar wind plasma, overcoming the limitations faced with
the legacy system. We also highlight some of the specific aspects of 
the OWFA, potentially relevant for the studies of coronal plasma 
and its turbulence characteristics.
\end{abstract}


\section{Introduction}

One of the successful observational programs being carried out for many
years at the Ooty Radio Telescope (ORT) is to determine the three-dimensional
structure, evolution and dynamics of quiet and transient solar wind in the
Sun-Earth space by using interplanetary scintillation (IPS). 
These studies have provided valuable inputs to the solar physics community 
to understand the physical processes in the heliosphere and space
weather events, such as coronal mass ejections (CMEs), interaction 
regions and their propagation effects. These studies have also provided a 
basis to develop models to predict the effects of such events at 1 AU
(e.g., \citealt{mano2006SoPh, vranak2008}).
However, the present sampling of the heliosphere by the ORT in a day limits 
the accuracy of such models. It is important to improve on this situation
because the practical consequences of space weather demand advance and 
accurate prediction of space weather effects associated with energetic CMEs.
For example, large and high speed Earth-directed CMEs are now considered to 
be the major source of transient interplanetary disturbances and shocks, 
which could cause severe geo-magnetic storms and lead to adverse effects on 
many crucial technological systems
(e.g., \citealt{manopankajwahab2011SoPh}).

A recent important concern of the solar physics is that a wide array of 
ground- and space-based observations show that the Sun has entered a 
sustained period of low activity (e.g., \citealt{manoapj2012, basu2016}).
The current solar cycle \#24 seems to be the smallest sunspot cycle 
observed since cycle \#14, which was observed around the beginning of 
year 1906. The effects of low magnetic activity are observed throughout 
the heliosphere. Among the prominent effects of this is that the cosmic 
ray flux near the Earth reached the highest level in the recent past. 
This in turn leads to various effects in the Earth's atmosphere. The 
state of the interplanetary medium associated with the low solar activity 
may also influence the effects of propagation of CMEs, formation of shocks, 
etc., (e.g., \citealt{mano2016ASPC}). 
Thus, an improved global monitoring of the solar wind and transients as well as 
the space weather conditions at the near-Earth distances would be highly timely.

\section{Solar Wind Measurements}

Solar wind observations, to date, have been made mainly in two ways, viz.: 
(1) directly by {\it 
in-situ} sampling of the downstream plasma, (2) remotely, through radio 
spectroscopy, and radio sounding observations. Since the 1960s a large number 
of spacecraft have successfully conducted routine {\it in-situ} sampling of
the solar wind (e.g., speed, density, temperature, composition, and magnetic 
field) in the near-Earth space and beyond. However, {\it in-situ} measurements
are limited to the one-dimensional time scan at the location of the spacecraft.
Proper measurements of spatial and temporal variations would require a grid of
space missions. Moreover, spacecraft sampling is largely confined to the 
ecliptic plane. Ulysses was the first and only space mission that probed the 
high latitude heliosphere (e.g., \citealt{GRL:GRL24924}).

However, the typical flow of the solar wind and the propagation of solar-generated 
transients show considerable changes with heliolatitude as well as
distance from the Sun. For example, the eruption of a CME/flare event, 
originating from the complex magnetic region on the Sun, adds significant 
amount of mass and magnetic field to the solar wind. The propagating 
transient associated with the CME is about a solar radius in size at the 
near-Sun region and it can expand to a structure of about an AU in the 
near-Earth environment (e.g., \citealt{mano2000ApJ}). 
Additionally, CME associated transients can involve: (1) acceleration of 
electrons and ions to high energies, (2) formation and steepening of 
interplanetary forward/reverse shocks at distances from the Sun, and 
(3) propagating/expanding large-scale magnetic clouds (i.e., flux ropes) 
(e.g., \citealt{mano2006SoPh,manoagalya2011advgeo}).
Thus, a high occurrence rate of intense flare/CME events can result in
the complex flow of solar wind of different time and spatial scales. 
Therefore, the study of global properties of the solar wind and the 
efficient tracking of a transient event (as well as in understanding its 
interaction with the ambient solar wind as a function of distance from Sun)
demand solar wind observations at consecutive parts of the heliosphere and 
we need to employ techniques and/or observations other than that of 
{\it in-situ} measurements from a satellite, which is normally confined 
to the orbit of the Earth. The radio remote-sensing technique, e.g., 
interplanetary scintillation, can provide estimates of solar wind 
speed and density turbulence in the three-dimensional space. However,
remote-sensing measurements probe the solar wind plasma integrated along
the line of sight and hence require care in the interpretation.

\subsection{Interplanetary Scintillation Technique}

The IPS technique exploits scattering of radiation 
from compact radio sources (i.e., radio quasars and galaxies of angular size,
$\Theta$ $<$500 milliarcsecond (mas)) by small-scale ($<$1000 km) density 
inhomogeneities in the solar wind (e.g., \citealt{hewish1964, coles1978, mano1993SoPh}).  
Since the drift of solar wind density inhomogeneities across the line of 
sight to the source causes intensity scintillation (typically, 
intensity fluctuations are observed on timescales of  $\sim$0.1--10 seconds),
the shape of the IPS power spectrum is related to the bulk velocity of the 
density structures and their scale sizes. 
The power spectrum of intensity fluctuations can be suitably calibrated to 
estimate the solar wind speed, the level of density turbulence, and the dominant 
scales present at the region of closest solar approach of the line of sight 
to the radio source (e.g., \citealt{mano1993SoPh, manoapj2012, mano1990, 
mano2000ApJ, tokumaru1994}).
Thus, the regular monitoring of the IPS of a given radio source using a 
single-antenna system of good sensitivity can provide the speed and density 
turbulence of the solar wind at a range of heliocentric distances (i.e., as the
source gradually approaches and recedes the Sun). Additionally, using IPS one can
do day-to-day monitoring of the heliosphere 
on a grid of large number of radio sources, whose lines of sight cut across 
different parts of the heliosphere. The Ooty Radio Telescope is well suited for 
such monitoring of the inner heliosphere.

\subsection{IPS with the Ooty Radio Telescope}

The Ooty Radio Telescope (ORT) is an equatorially mounted steerable cylindrical 
paraboloid (530 m$\times$30 m) (\citealt{swarup71}), operating at a central 
frequency of 326.5 MHz. The legacy system of the ORT provides a resolution 
of 2$^\circ\times 6'$, in the east-west and north-south directions, 
respectively. The ORT covers $-4^{\rm h} 7^{\rm m}$ to $+5^{\rm h} 25^{\rm m}$ 
in hour  angle and $\pm$65$^\circ$ in declination. The ORT is also 
supported with a beam-forming system of 12 simultaneous equispaced beams. Each 
beam is separated by 3$'$. For a given declination pointing, this system provides 
a sky coverage of 36$'$ in the north-south direction 
(\citealt{sarma1975, selvanayagam93}).
Over the last several years, IPS observations made with the ORT have been 
efficiently used to study the dynamics of the interplanetary medium on 
time scales ranging from a few hours to years. 

In the IPS observing mode,
since the ORT is fully steerable in the east-west direction, a given scintillating
source can be continuously tracked for $\sim$09$^{\rm h}$ 30$^{\rm m}$ (i.e., from 
the rise time of the source at the east limit of the ORT to its set time at the 
west limit) and the solar wind condition along the direction of the source can be 
monitored in details. In the above span of observing time, the change in the 
position of the source with respect to the Sun is not significant (i.e., typically
a source moves with respect to the Sun by $\sim$1 degree per day) and the continuous
tracking observation essentially provides the parameters of the solar wind along a 
fixed line of sight. On the other hand, to scan the solar wind over a wide area of 
the sky, the ORT can be parked at a convenient hour angle and each source transiting 
at the telescope can be observed for about 90 seconds by just electronically switching
the north-south beam. A sufficient time gap, if available between the observations of 
two consecutive sources, can be used to observe the off-source level of IPS, which is 
required for the estimation of the scintillation index of each observed source
(\citealt{mano1993SoPh, mano1995SoPh}).
In a regular mode of solar wind monitoring, a parking hour angle is chosen close to the 
east limit of ORT ($-4^{\rm h}$), which can include the global monitoring of regions,
respectively, in the western side of the Sun, close to the Sun, and in the eastern 
side of the Sun. In such an observing mode, depending on the duration of observation 
(e.g., 6--8 hours), the scintillating sources monitored can cover a wide range of 
heliolatitudes as well as heliocentric distances in the range of 20--250 {\rsun} in
the west as well as east of the Sun and a CME and its associated disturbances 
propagating in any direction with respect to the Sun can be recorded. Further, the 
steerability of the ORT allows us to monitor the same part of the sky by parking
the ORT, respectively, at hour angles overhead ($0^{\rm h}$) and near the west limit 
($+4^{\rm h}$) and provides the opportunity to study the temporal and radial evolution
of the CME for a period of $\sim$20 hours (e.g., \citealt{mano2000ApJ, mano2001ApJ}).
Alternatively, if an {\it a priori} information on the propagation direction of a 
CME is know, the ORT can also be parked at an appropriate hour-angle position and 
the scintillation of the drifting sources can be studied for a duration of up to 
$\sim$20 hours, which can also cover a wide heliocentric distances in the west and 
east of the Sun. 

Moreover, for a given central beam setting of a radio source, the 12-beam 
system of the ORT often includes one or more sources at other 
beams away from the central beam and such observations increase the number 
of sources monitored at a given time. By combining the above hour angle 
scans, a good spatial resolution in the sky plane is achieved. In the recent 
past, depending on the other observing programs, the ORT has been used to 
make daily observations of about 600--900 radio sources, covering a distance 
range $\sim$20--250 {\rsun} (corresponding to a solar elongation range of 
$\sim$5$^\circ$--100$^\circ$) in the western and eastern sides of the Sun. 
The above set of sources monitored normally includes nearly $\sim$90\% of 
scintillators and the remaining are non-scintillators (sources having large 
angular sizes, $\Theta$ $>$500 mas). The extended sources are not influenced 
by the solar 
wind conditions and are monitored to check the contamination caused by the 
ionospheric plasma and the level of radio frequency interferences (RFIs), if
present. It is to be noted that the ionospheric scintillation, which is rare 
at 326.5 MHz during the daytime, has different timescales from that of IPS and 
can also be easily eliminated from the IPS power spectrum.

In the past, a good fraction of the ORT time was used for IPS observations.
The regular long-term monitoring of IPS on several radio sources has in fact 
provided the ``{\it scintillation index -- heliocentric distance}'' curves
of all these sources for several years. Ooty 
scintillation-index curves are available for $\sim$8000 radio sources and 
these can be used to estimate the typical one-dimensional angular diameter of 
the compact component of each source as well as its flux density at 326.5 MHz 
(Manoharan et al., {\it in preparation}). 
Additionally, the scintillation curves obtained from the ORT have been made
available to IPS groups at the Solar-Terrestrial Environment Laboratory, Nagoya 
University, Japan (\citealt{tokumaru2011}) 
and the Mexican IPS Radio Array at the University of Mexico, Mexico 
(\citealt{americo2004}).

Ooty IPS studies have led to several important results on -- (1) understanding 
otherwise unavailable information on the three-dimensional shape and dynamics 
of the large-scale structures of CME-associated disturbances in the Sun-Earth space
(e.g., \citealt{mano2010SoPh});
(2) clearly establishing  the speed and associated magnetic structure of CMEs 
as well as the background solar wind, which are crucial to understand the effects
of the CME at the near-Earth environment (\citealt{mano2001ApJ, mano2006SoPh}) 
and these studies have led to the development of propagation models to predict 
the effects of CME events at $\sim$1 AU (e.g., \citealt{model2015SoPh});
(3) the relationship between the speed and transit time of CME-associated 
disturbances; 
(4) the three-dimensional evolution of density turbulence and speed of 
quasi-stationary solar wind at various levels of solar activity over a 
period of three decades (e.g., \citealt{mano1993SoPh, manoapj2012}); 
(5) a better understanding of the nature of turbulence and plasma processes 
of the ambient solar wind in the inner heliosphere and the solar wind 
plasma associated with the interplanetary disturbance (e.g., shape of the 
density turbulence spectrum and its dissipative scales as the function 
of distance from the Sun) (\citealt{mano1994, mano2000ApJ, mano1987sowi}).

\begin{figure}
\centering
\includegraphics[width=.47\textwidth]{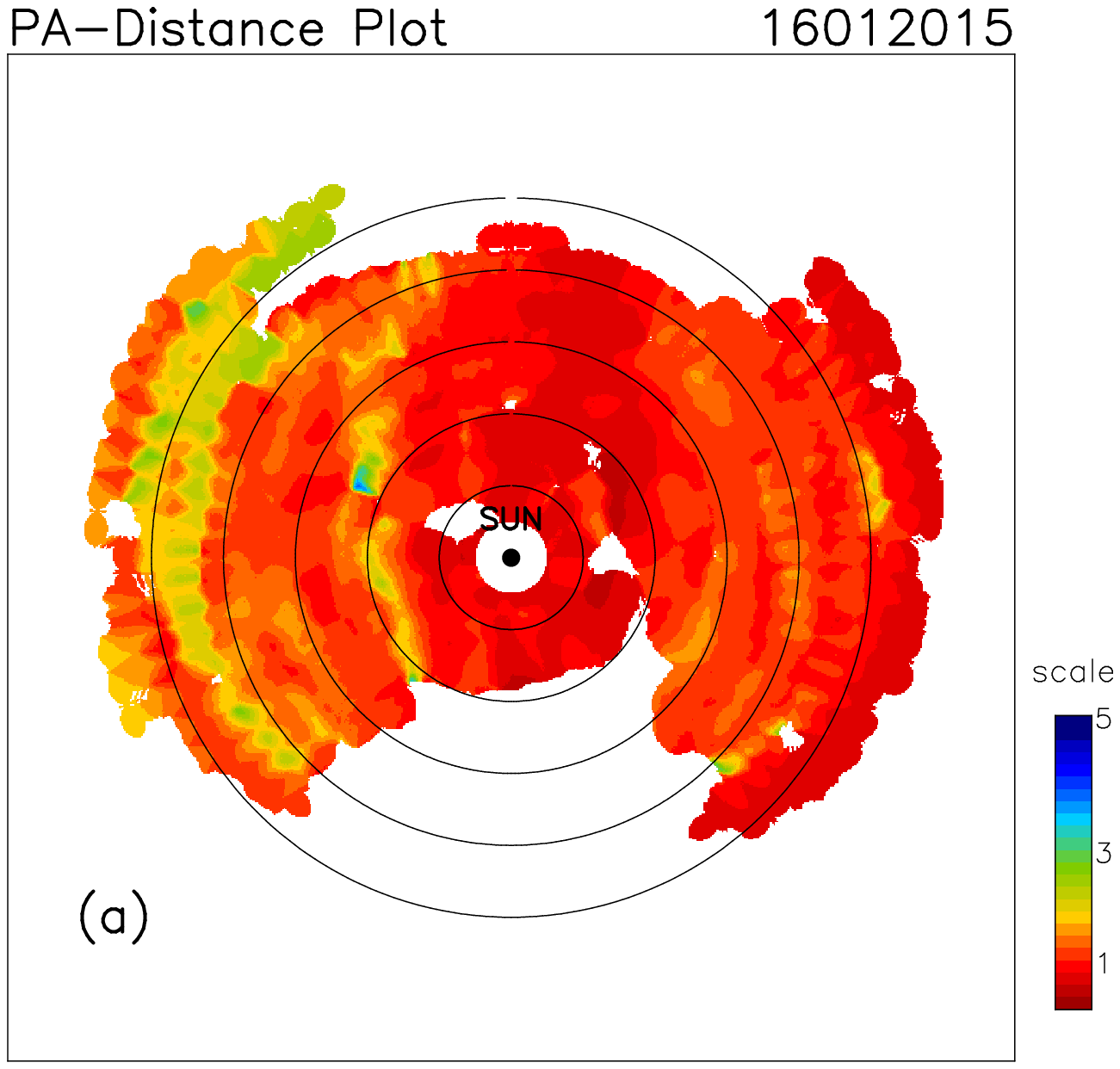}\hfill
\includegraphics[width=.43\textwidth]{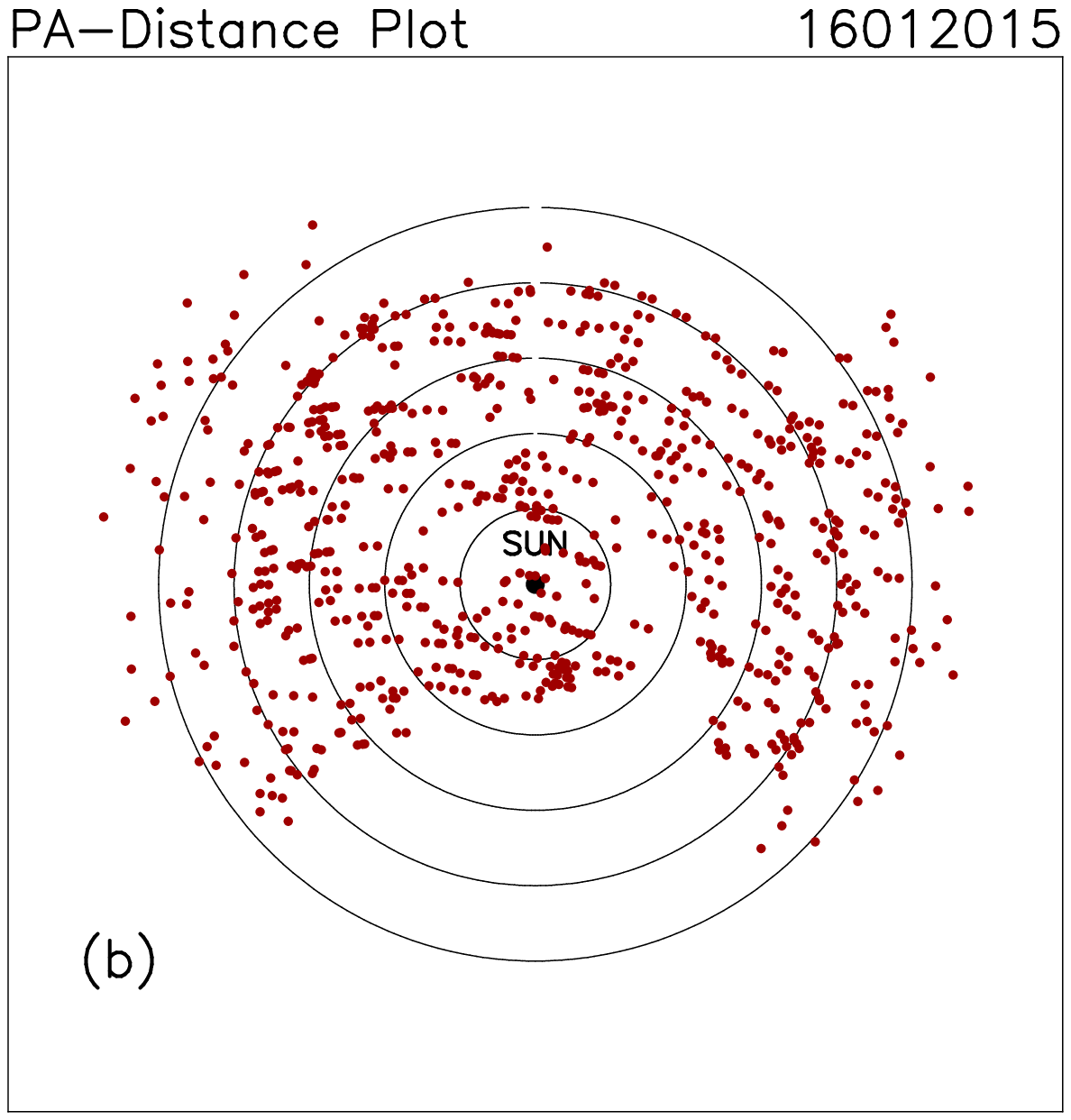}
\caption{ (a) An example of scintillation image observed with the ORT on 16 January 
2015 (left plot). It shows the sky-plane projection of observed levels of density 
turbulence in the solar wind in the inner heliosphere and the Sun is at the centre
of the image. The red color code indicates the background solar wind and the enhanced 
scintillation patches indicate the presence of interplanetary CMEs in the IPS field 
of view. 
(b) A similar plot shown in the right is the distribution of IPS sources observed 
to make the scintillation image and each dot represents the position of a radio 
source. The plotted IPS field of view includes $\sim$600 radio sources. In these  
``{\it position angle (PA) -- heliocentric distance}'' plots, the north is at the 
top, i.e. PA = 0$^\circ$ and PAs 90$^\circ$, 180$^\circ$, and 270$^\circ$, 
respectively, correspond to the east, south, and west of the Sun. Observing time 
increases from the right-hand side of the plot (i.e., from the west of the Sun) to 
the left-hand side (i.e., the east of the Sun). The concentric circles are of radii, 
50, 100, 150, 200, and 250 {\rsun}. }
\end{figure}

\subsection{IPS Imaging with the Legacy ORT System}

Figure 1a shows example of a scintillation image obtained with the ORT based on 
observations of about 600 IPS sources on 16 January 2015. These observations
were taken over a time period of $\sim$18 hours at an hour-angle pointing of $0^{\rm h}$
and nearly equal regions were sampled in the western and eastern hemispheres
of the heliosphere.  This plot is the sky-plane 
snapshot image of level of density turbulence in solar wind, observed over a
heliocentric distance range of more than an AU from the Sun. (For more details 
on the Ooty IPS analysis and the construction of scintillation image, the reader
may refer to \cite{mano2001ApJ}.) %Manoharan et al. 2001.)
The image area is made of 600$\times$600 pixels (i.e., covering a heliosphere 
diameter about 3 AU). Since the image is made from scintillation observations of
a large number of discrete radio sources, it has been smoothed by a two-dimensional 
Gaussian of width 20$\times$20 pixels.
The radius of the concentric circles drawn around the Sun increases in steps of 
50 {\rsun}. The orbit of the Earth is located at $\sim$215 {\rsun}.
The observing time increases from the right of the image (the west of the Sun) 
to the left (the east of the Sun). The red color code indicates the background 
(i.e., ambient) solar wind. The enhanced scintillation regions represent the 
CME-driven disturbances and/or interactions between slow and fast solar wind
streams. For example, the propagation of two CMEs that originated from the Sun 
on 13 and 15 January 2015 are seen in the north-east direction of the Sun, 
respectively, at $\sim$225 {\rsun} and 100 {\rsun}. Apart from these, some more
faint enhanced features are also identifiable in Figure 1a.

As mentioned above, this scintillation image has been made from IPS observations 
of $\sim$600 
radio sources and the distribution of these sources around the Sun is shown in 
Figure 1b (right-hand side plot). Each dot in the plot corresponds to a radio 
source. The noted asymmetry in the source distribution between the north and 
south regions of the heliosphere is due to the allowed declination coverage of 
$\pm$60$^\circ$ for the above IPS measurements (i.e., at high declinations,
$>|\pm45^\circ|$, the sensitivity of the ORT is severely affected by the 
decrease in the projected collecting area).
The radio sources employed are of flux density $\geq$1 Jy at 326.5 MHz. It may 
be noted that the overall source distribution is not uniform and there are 
several small to large gaps seen in the plot. In particular at heliocentric 
distances $>$150 {\rsun}, the number of sources observed is less than 
that observed at small distances from the Sun. This is mainly due to the fact 
that the radial falloff of the power of density turbulence in the solar wind 
is rather steep, $C_{N_e}^{2}(R) \sim R^{-4}$, and the scintillation of a 
source would also decrease with the distance from the Sun in the above manner. 
At 326.5 MHz, for a point source (e.g., $\Theta$ $\leq$20 mas), the 
maximum level of scintillation is observed at $\sim$40 {\rsun} and it is 
reduced by an order of magnitude at $\sim$150 {\rsun} (\citealt{mano1995SoPh}).
In the present legacy system of the ORT, we employ a receiver system bandwidth 
of $\sim$4 MHz centered around 326.5 MHz and use an integration time on each 
source of $\sim$90 
seconds. This enables the detection of a scintillation flux of $\sim$25 
mJy. In the legacy system, for a given observing period, the number of sources
observed is therefore limited by their compact component flux densities as 
well as angular sizes.  
However, the Gaussian smoothing function used to smooth an image area of 
$\sim$20$\times$20 pixels enables the detection of overall large-scale 
structures of the three-dimensional solar wind. Further, CMEs in general 
propagate with high speed (i.e., speed in excess of the background solar 
wind speed of $\sim$300-350 {\kmps}) and the present spatial resolution 
provided by the distribution of sources is just enough to sample the 
propagating CME structures at a couple of places in the inner 
heliosphere (e.g., \citealt{mano2001ApJ}). 
Thus, IPS observations with the present legacy system have 
limited the detailed inference of the small-scale features of solar wind, 
CMEs, and interaction regions. Another important point is that the temporal 
cadence obtained for a fast moving CME in a day is low and hence
precise study of such fast moving CMEs is difficult with the present IPS 
measurements alone. 

\section{IPS with the Ooty Wide Field Array (OWFA)}

In order to improve the study of fine features of CMEs, solar wind structures
and to predict their arrival times and associated consequences at the near-Earth 
space about a day in advance, the essential requirement is the imaging of the 
interplanetary medium with further increased spatial and temporal resolutions. 
In comparison with the legacy ORT system, the upgraded OWFA has both a large 
instantaneous bandwidth of $\sim$38 MHz and a significantly wide north-south beam of 
$\sim$27$^\circ$ (refer to Subrahmanya, Manoharan, and Chengalur 2016, {\it in this volume}). 
For example, at a given time, the large declination coverage of the OWFA beam 
would allow the simultaneous observation of a number of IPS sources covered
within its beam. Additionally, the increase in the sensitivity of the OWFA by
a factor $\sim$3 would enable the observation of sources with weak scintillating
flux density. 

The OWFA will have an FX correlator with a bandwidth of $\sim$38~MHz 
and with 800 spectral channels (Subrahmanya, Manoharan \& Chengalur, 2016,
JApA ({\it submitted}); Subrahmanya \& Chengalur 2016 ({\it in preparation})).
Phased array beam formation will be done post-correlation. Since the OWFA
baselines are highly redundant, (with only 264 of the 34716 baselines
being unique) the post correlation beam formation can be done highly
efficiently. The visibility on the each of the unique baselines can
be estimated by averaging over the multiple redundant copies. The data
volume is now so drastically reduced that one can form simultaneous
phased array beams covering the entire field of view corresponding
to the primary beam of the individual OWFA elements. All sources of
interest lying within the primary beam can hence be simultaneously 
monitored. This real-time beam forming requires high cadence, real-time 
calibration of the antenna gains, for which fast redundancy 
calibration algorithms have been developed and tested for OWFA 
(\citealt{marthi2014MNRAS}).

\begin{figure}
\centering
\includegraphics[width=.43\textwidth]{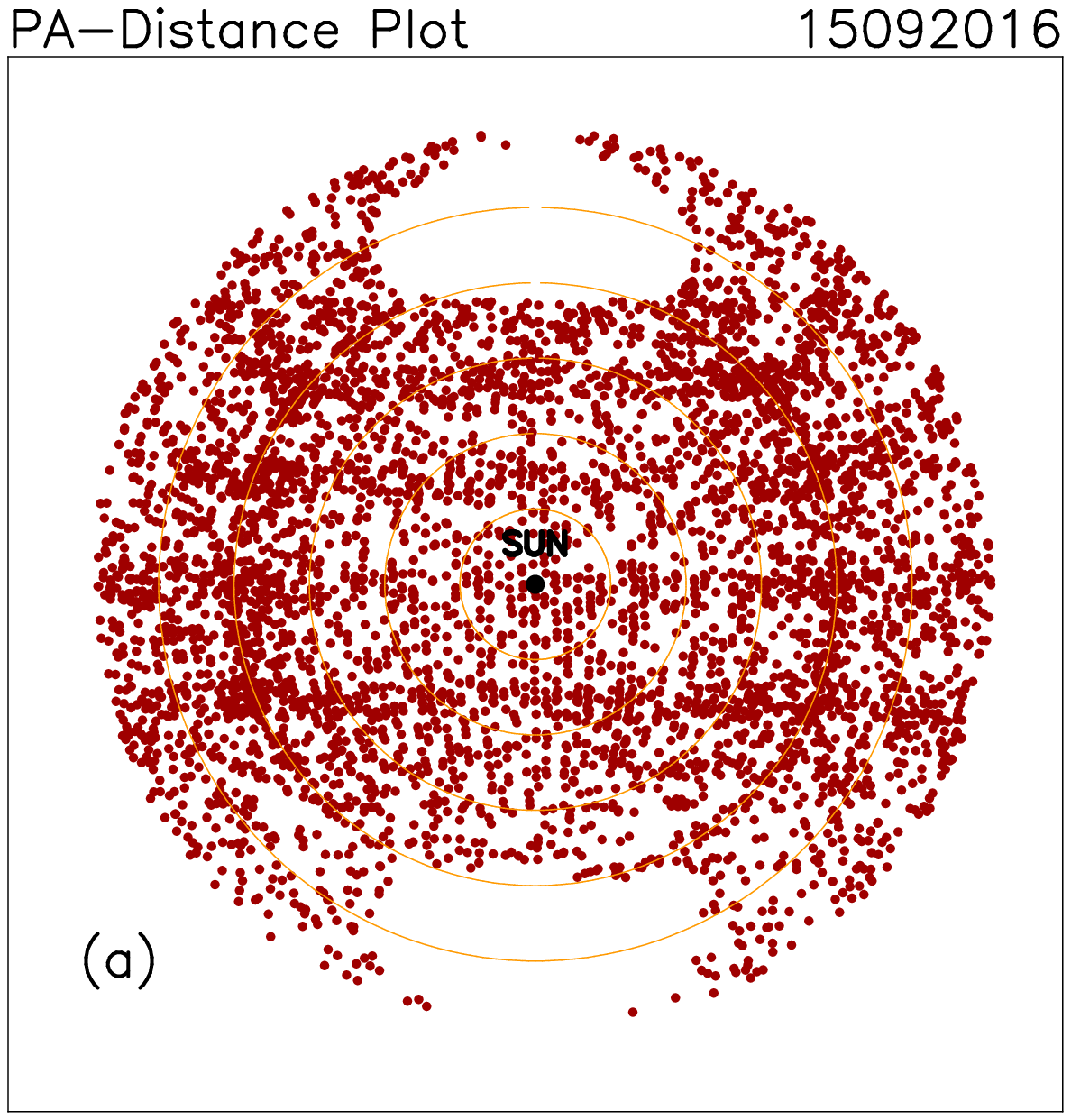}
\includegraphics[width=.47\textwidth]{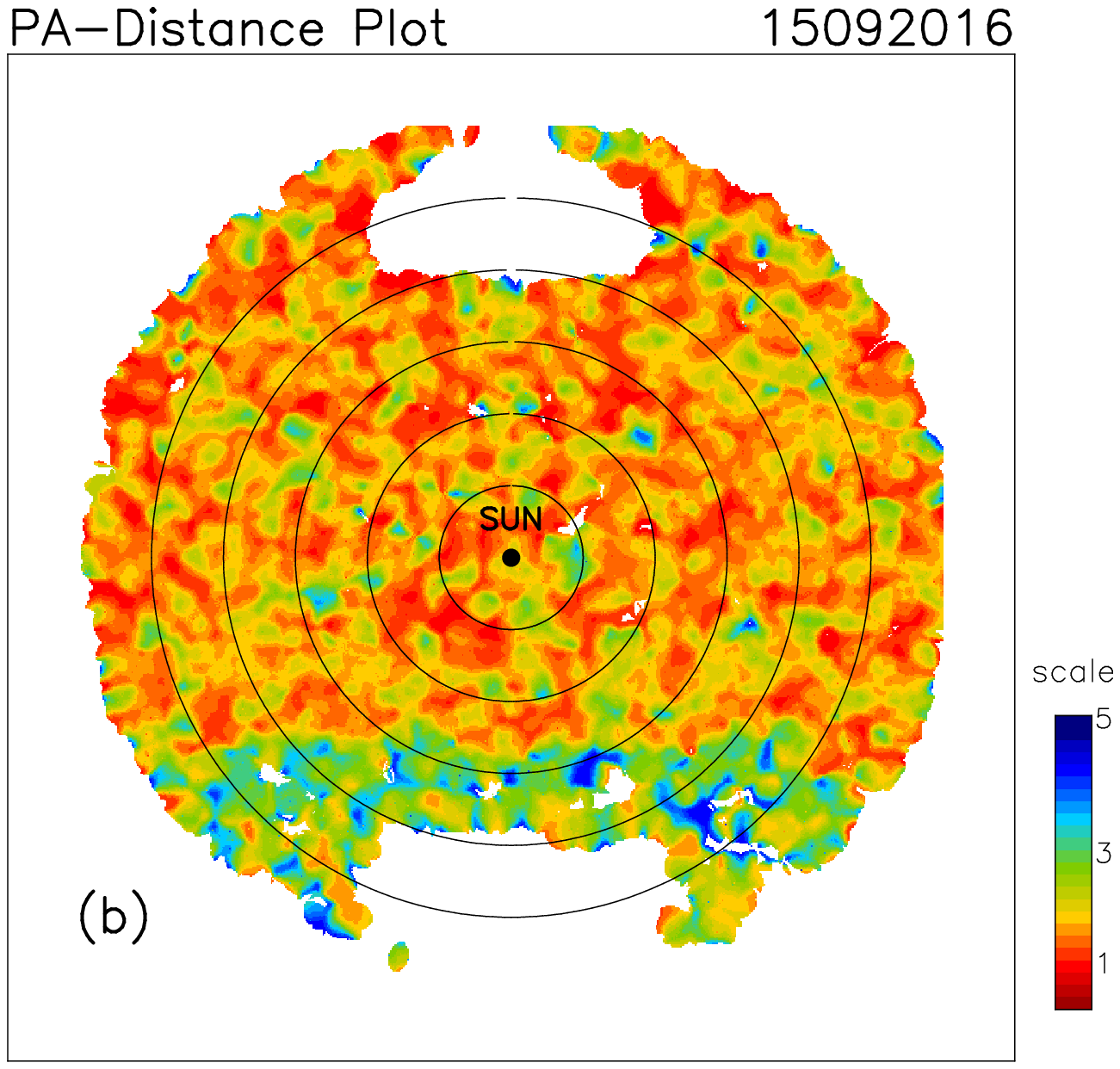}
\caption{(a) A simulated source distribution possible with the OWFA 
for an IPS monitoring on 15 September 2016. It includes $\sim$5500 
radio sources and each dot on the plot corresponds to a radio source.
(b) The simulated scintillation image obtained for the above source
distribution. These plots are similar to the format of Figure 1.
}
\end{figure}

For the parameters of the OWFA system, a simulated source distribution 
(for observations on 15 September 2016) is shown
in Figure 2a. The source lists used for making the above simulation include
(1) the existing list of IPS scintillators obtained from the ORT long-term 
monitoring, which includes sources of flux density $\geq$1 Jy at 326.5 MHz, 
and (2) sources of flux density $\geq$0.7 Jy at 326.5 MHz, scaled from the 
Texas and Molonglo surveys, respectively, made at 365 and 408 MHz
(\citealt{texas1996, molo1981}).
This simulation has been done for an hour-angle pointing of $\sim$0$^{\rm h}$ 
and an observing time span of $\sim$18 hours, centered around the transit 
time of the Sun on 15 September 2016 and it is generated for an observing 
situation similar to that presented in Figure 1. The OWFA beam has been 
scanned from $-65^\circ$ to $+65^\circ$ in a step of $\sim$20$^\circ$, 
which allows considerable overlap between one beam setting to the other. In a beam 
setting, all the sources of flux density $\geq$0.7 Jy contained in the 
primary beam area of $2^\circ \times 27^\circ$ are considered. However, 
if more than one source is located within the beam size of the full ORT,
$2^\circ \times 6'$, it is counted as a single source. The number of sources
included in the simulated distribution is more than 5500, which is about an
order of magnitude larger than the number of sources observed with the legacy
system (refer to Figure 1). In comparison with the legacy system, the above
source distribution obtained for the OWFA nearly fills the imaged area and 
large gaps are almost absent. Since the improved sensitivity of the OWFA allows 
the observations of sources with small scintillating flux density, the 
simulated coverage looks nearly uniform  throughout the 3-AU heliosphere. 
It is to be noted that the above simulation 
includes only sources of S$_{\rm 326.5}  \geq$0.7 Jy. But, since IPS is not 
much limited by the confusion limit of the system, the OWFA can therefore 
observe sources weaker than the above limit. Since the number density 
of the source increases with decreasing flux density, the IPS with the 
OWFA is expected to provide more finer spatial resolution than that 
discussed above.

Figure 2b shows the typical two-dimensional image obtained from the above
simulated source distribution. In order to make this image, a uniform 
level of 50\% of the source flux density is taken as the level of 
scintillation and the maximum allowed scintillation is limited at 5 units.
It is equivalent to Figure 1a and covers an area of 600$\times$600 pixels. 
The simulated image has been smoothed by a two-dimensional Gaussian of width 
5$\times$5 pixels, which is 16 times smaller than the area used in smoothing
the image obtained with the legacy system (refer to Figure 1).
It is evident that the image is not smooth and shows number of fine 
features. Such an image on a short observing time span will provide 
information on small-scale structures
embedded within a CME as well as their evolution characteristics. The
quick scanning of the OWFA beam would also allow to image the fast 
moving CMEs and study their dynamics with the IPS measurements alone. 

IPS observations of small flux density sources aimed with the OWFA are 
also likely to lead to the detection of ultra compact radio sources 
possibly at high red shift. 
The increase in sensitivity of the OWFA by a factor of $\sim$3 not only 
enables the observation of small flux density sources, but will also 
provide an excess of $\sim$5 dB signal in the estimated power spectrum 
(e.g., \citealt{arun2015}).
This will be useful to infer the characteristics of small dissipative 
scales (i.e., at scales $<$25 km (\citealt{mano1994, mano2000ApJ, mano1987sowi}))
of the density turbulence. These
studies are crucial to understand the evolution of density turbulence 
and the behaviour of the dissipative scales in the solar wind, which,
in turn are essential to understand the physical processes associated 
with the solar wind plasma. The turbulence spectrum can also be 
studied as functions of (1) distance from the Sun, (2) heliographic 
latitude or source region of solar eruption, and (3) solar activity 
phase.

\subsection{Reconstruction of Heliosphere}

Since the IPS with the legacy system provides observations on a grid of 
large number of sources per day, it allows the reconstruction of the 
three-dimensional heliosphere using the time-dependent computer-assisted 
tomography technique (CAT) developed at the University of California, 
San Diego (e.g., \citealt{mano2010SoPh, jackson2003}).
The basic data sets required for the three-dimensional reconstruction 
are the time series of solar wind speed and level of scintillation for 
a large number of lines of sight of the heliosphere. 
%For these studies, 
%when both speed and scintillation are available for a given line of 
%sight, they can be considered as parameters for the reconstruction. 
Even with the limited number of lines of sight, the reconstructions made 
with the legacy system of the ORT have given several important results
(\citealt{bisi2009, mano2010SoPh}). 
The much better coverage of the OWFA would hence provide an improved 
understanding of the dynamic heliosphere.

\subsection{IPS Picket Fencing with the OWFA}

In order to probe fine structure in a CME, the legacy ORT system was
employed in a couple of special type of IPS observations, in which only 
a few IPS sources (i.e., only two or three lines of sight) were 
selected such that to lie at the expected crossing location of the 
radially moving CME and its associated interplanetary disturbance.
The continuous monitoring of a few radio sources
at frequent interval (about 2 min on each line of sight) allows one
to estimate (1) ambient solar wind before the 
arrival of the CME at the IPS lines of sight; (2) the properties of 
plasma along the radial cut through the shock, sheath, and CME; (3) 
the upstream flow behind the interplanetary disturbance
(\citealt{mano2010SoPh}). 

Even though IPS measurements are integrated along a line of 
sight, they are still useful to study the characteristics of the 
plasma within the CME and the solar wind moving ahead and behind 
the disturbance.
The CME study using this `picket-fence' method with the OWFA will allow
observations of large number of lines of sight for a given distance
from the Sun. For example, in Figure 2a, if we consider a circular 
ring of constant radius running from the north pole to south pole 
of the inner heliosphere,
several sources will lie on the circular ring. The simple beam
switching between these sources would provide high cadence sampling,
at times before, during a CME passage, and after its crossing. Such
set of observations at two or more radial distances can give detailed 
information on (1) internal structure of the CME and its variations as
a function of radial distance, (2) overall size of the CME and its
evolution with distance, (3) the shock forming location in front of 
the CME, (3) stand-off distance/time (i.e., distance or time between
the driver gas and the shock), and (4) dynamics of the shock and sheath. 
Such studies can also be useful to infer the particle acceleration 
site as well as the effective acceleration encountered by the CME in 
the solar wind (\citealt{mano2006SoPh}).
They are also vital to understand the 
energy of the solar wind as well as the energy transfer between the 
disturbance and the surrounding ambient medium.

As in the picket-fence method, if we examine the circular
rings of varying radii at the polar regions above the Sun, it can
provide new information on the fine structure of polar plumes, 
their temporal evolution and relationship with other transient 
phenomena. Polar plumes are thin long ray kind of 
structures observed above the polar coronal holes and they can
extend up to a distance of several solar radii. The OWFA would help to
observe the polar solar wind on much finer spatial and temporal
scales, allowing the study of plasma processes associated with
the polar plumes. In a similar way, OWFA IPS program will also 
help to investigate the plasma associated with coronal jets 
above the limb of the Sun. Such studies could be essential 
in understanding the plasma processes linked to the coronal 
heating problem. 

\section{Summary}

In this paper, we briefly discuss the ongoing IPS observing program with
the legacy system of the ORT and its limitations in understanding the 
details of physical processes associated with CMEs and the solar wind
in the Sun-Earth distance range. The upgraded ORT, the OWFA, will have 
a large field of view and improved sensitivity and allow observations 
of about an order of magnitude large number of IPS sources. 
Thus, its unique ability to sample large areas of the interplanetary 
medium with unprecedented spatial and temporal resolutions gives the 
OWFA advantages over all current radio telescopes.
The OWFA can be potentially useful in addressing, but not limited to, 
the following studies: (1) it can 
enhance the inference of dynamics of space weather events, in particular,
fast moving CMEs and improve the predictability of their arrival at the 
near-Earth space; (2) The tomographic reconstruction of IPS with OWFA is 
expected to provide a clear view of the global heliosphere; (3) Detailed 
IPS studies on limited number of radio sources can be useful to probe
different parts of a CME, its associated shock, sheath, and their influence
on the surrounding ambient solar wind; (4) The capability of OWFA also allows
to select a given region of the heliosphere and investigate its plasma
characteristics.

\section*{Acknowledgment}
We thank the observing/engineering team of Radio Astronomy Centre for 
the help in forming the critical component of this project. We thank
A. Johri for helpful conversation and reading of the manuscript. This 
project is partially supported by the Indian Space Research Organisation
(ISRO sanction order no. {E.33011/35/2011-V}).

\bibliography{paper_18aug2016}{}
\bibliographystyle{plainnat}

\end{document}